\documentclass[useAMS,usenatbib]{mn2e}
\usepackage{times, epsfig}

\def\ten#1{\times 10^{#1}}
\def\Amax{A_{\rm max}}
\def\umin{u_{\rm min}}
\def\that{{\widehat t}\,} 
\def\etal{et al.\ }

\title[Microlensing puzzle]{Contamination in the MACHO dataset and the puzzle of LMC Microlensing}
\author[Griest \& Thomas]{Kim Griest\thanks{E-mail: griest, clt@ucsd.edu} and Christian L. Thomas\footnotemark[1]\\
Physics Department, University of California, San Diego, La Jolla, CA  92093}

\begin{document}
\date{draft: 1/14/05}
\pagerange{\pageref{firstpage}--\pageref{lastpage}} \pubyear{2005}

\maketitle

\label{firstpage}

\begin{abstract}
In a recent series of three papers, Belokurov, Evans, \& Le Du (2002, 2004), 
and Evans \& Belokurov (2004),
reanalysed the MACHO collaboration data and gave alternative
sets of microlensing events and an alternative optical depth to 
microlensing toward the Large Magellanic Cloud (LMC).
Even though they examined less than 0.2\% of the data they claimed
that by using a neural net program they had reliably selected 
a better (and smaller)
set of microlensing candidates.  Estimating the optical depth from this
smaller set, they claim that the MACHO collaboration overestimated the optical
depth by a significant factor and that the MACHO microlensing experiment
is consistent with lensing by known stars in the Milky Way and LMC.
As we show below, the analysis by these authors contains several errors
which render their conclusions meaningless.  Their efficiency
analysis is clearly in error, and since they did not search through the
entire MACHO dataset,
they do not know how many microlensing events their neural net
would find in the data or what optical depth their method would give.
Examination of their selected events suggests that their method 
misses low S/N events and thus would have lower efficiency
than the MACHO selection criteria.  In addition, their method
is likely to give many more false positives (non-lensing events identified
as lensing).  Both effects would increase their estimated optical depth.
Finally, we note that the EROS discovery that LMC event-23 is a variable
star reduces the MACHO collaboration estimates of optical depth and Macho
halo fraction by around 8\%, and does open the question of additional
contamination.
\end{abstract}


\section{Introduction}
A number of experiments have returned microlensing results on the nature of
the dark matter in the Milky Way.  
The EROS collaboration (Aubourg, \etal\ 1993; Lasserre \etal\ 2000) monitored 
17.5 million stars in the LMC for 2 years and 5.3 million stars for 6 years
finding 3 microlensing
candidates.  Using these events and an efficiency analysis they found an upper limit of 40\% (95\% C.L.) on the halo mass fraction in Machos in the mass range of $10^{-7}M_{\odot}$ to $1M_{\odot}$.  The MACHO 
collaboration (Alcock, \etal 1993; 1997; 2000)
monitored 11.9 million stars in the LMC over 5.7 years, finding
between 13 and 17 microlensing candidates.   They did a careful
efficiency analysis, including the effect of blending in their crowded fields
and found a best fit Macho halo fraction of 20\% 
($0.08 < f <  0.5$ at 95\% CL), which
corresponds to an optical depth of $\tau = 1.2^{+0.4}_{-0.3} \ten{-7}$.  
It is important to realize that the MACHO estimate of 
dark matter halo fraction depended
crucially on the use of standard models of the Milky Way thin disc, thick
disc, spheroid, halo, and the LMC disc and halo.  
In particular they estimated that between 2 and 4 events ($\tau$ between
$0.24 \ten{-7}$ and $0.36 \ten{-7}$.
should be from lensing of known stars, depending on Galactic and LMC structure.
Several groups (e.g. Wu 1994; Sahu 1994; Gates, \etal 1996; Evans, \etal 1998; Gould 1998;
Aubourg, \etal 1999; Alves \& Nelson 2000; Weinberg 2000; Evans \& Kerins 2000; DiStefano 2000;
Gyuk, \etal 2000; Mancini et al. 2004)
have investigated whether additional 
stars in the LMC, or modification to the Milky Way structure could explain the
extra microlensing events, with the augument being made both ways.
If the extra microlensing events are due to a new halo population of 
Machos (white dwarfs? primordial black holes?),
they would represent roughly the mass of the Milky Way disc and would have
a major effect on Galactic chemical evolution, galaxy formation and 
cosmology.  Thus, currently the most popular hypothesis is probably that 
the LMC has an additional, as yet undiscovered\footnote{See Minitti, et al. 2003 for recent developments.}, extended population of stars,
that is producing the extra microlensing.  The mystery of the extra 
LMC microlensing has thus been phrased as the question of whether there is LMC 
self-lensing or a new population of halo objects that make a small but 
significant contribution to the dark matter.

Belokurov, Evans, \& Le Du (2003, 2004) , and 
Evans \& Belokurov (2004) (BEL/EB) 
have suggested instead that the MACHO collaboration sample of microlensing
events is severely contaminated with non-microlensing events, 
and that our optical
depth is thus substantially over-estimated.  
This is certainly a possibility as has
been known from the beginning of the microlensing surveys, and is a topic
worthy of investigation. 
BEL/EB reanalysed 22000 random MACHO light curves and 29 MACHO microlensing and 
SNe candidates using a neural net and selected 7 microlensing 
events (BEL 2004).  In their first paper on the LMC, BEL
recognized that they
had not analyzed all 11.9 million light curves, and
did not claim to prove that the MACHO
optical depth was greatly overestimated, but did strongly suggest that 
that was the case (BEL 2004).
In talks, and in the most recent paper 
(Evans \& Belokurov 2004), they were less cautious; in EB 2004
an efficiency calculation was performed and 
an optical depth was calculated. 
Strong claims are now being made that contamination in the MACHO data set 
is the solution to the mystery of LMC microlensing (EB 2004).
However, as we show below, there are severe errors in the BEL/EB analysis,
and so it is not yet useful in determining whether the MACHO optical depth 
is over-estimated.

Fortunately there are several upcoming experiments that can address and 
hopefully solve the puzzle of the extra LMC microlensing.  
The MEGA (de Jong \etal 2004) and 
POINT-AGAPE (Paulin-Henriksson \etal 2003) experiments are monitoring
stars in M31.  This is hopeful, since the large difference in optical depth
between sources on M31's near side and far side can distinguish between
halo and disc microlensing.  Preliminary results are still ambiguous
with MEGA finding weak evidence in favour of halo microlensing with 4 events (3 on the far side and 1 on the near side), and POINT-AGAPE finding weak evidence against halo microlensing with 3 events.  In addition, the SuperMACHO collaboration (Becker \etal 2004) hopes to find
enough new LMC events to clarify the situation.  The ultimate test may
come from space missions, for example,
the Space Interferometry Mission microlensing key project (Unwin \& Turyshev 2002)
which can measure the distance to the lenses (Gould \& Salim 1999).
Even determining the distance
to 3 or 4 lenses will conclusively solve the problem of whether the extra
microlensing is due to LMC or Halo lenses.  In addition there is a proposed
Deep Impact extended space mission (DIME) (Cook, \etal 2003) which is dedicated
to microlensing parallax measurements from space and could 
answer the question if approved.

Since these experiments are still underway, it
is important that the community knows that the puzzle
has not yet been solved and that the BEL/EB claims should be
treated with caution.

\section{Event selection and efficiency calculation}

The optical depth to microlensing is just the chance that a given
source star is undergoing microlensing.  Since this probability is small,
one monitors a large number of stars, $N$, for a long period of time, $T$, and 
estimates the optical depth by
\begin{equation}
\tau = {\frac{\pi}{4 N T}} \sum_{events} \frac{\that_i}{\epsilon_i},
\end{equation}
where the duration of the event, $\that$, is the Einstein ring diameter crossing time
and $\epsilon_i$ is the efficiency for that event (probability of 
the combination of the experiment
and selection criteria finding an event with duration $\that$ in
that field.)

In order to estimate $\tau$ one needs to search through all $N$ light curves
and use some automated selection
criteria to pick out the microlensing candidates.  This gives the set
of microlensing events and their durations.  Then one must apply 
exactly the same analysis and selection criteria to a large 
sample of simulated microlensing light curves and determine $\epsilon_i$ 
as the fraction of simulated events that the analysis plus selection criteria 
find.

Every set of selection criteria will miss some real microlensing events, and 
also identify some non-microlensing events as microlensing, so developing
selection criteria is a trade-off between losing real microlensing events
and contaminating the sample with non-microlensing.  
Allowing false positives erroneously increases the optical depth, 
while cuts that are too restrictive eliminate actual events resulting 
in larger errors on the measured optical depth.  
Contamination is more serious than missing events since, if properly calculated,
a decrease in efficiency will compensate for the reduction 
in number of events.
Thus ideally, one uses the broadest set of cuts that appears to give
negligible contamination by false positives.

In the above framework,
a neural network program is just a set of selection criteria, and so the above
procedure applies.
Neural network programs have found wide use recently for data analysis
in particle physics, astrophysics, and many other arenas.  They have been
found to be good at making efficient cuts in multi-dimensional parameter
spaces, so we agree that the main
idea of BEL/EB, redoing the MACHO event selection with a neural net, 
is a good one.
However, any set of selection criteria, even a neural net, is only as good
as the data input to it and the way it is used to calculate the efficiencies
and the final results.

So, first the raw light curve data must be converted into statistics
that capture all the essential distinguishing properties of the microlensing
events and the background.
Then, after training the neural net on simulated events and some real and/or
simulated background, all $N$ light curves should be fed to the net so
it can produce a set of microlensing candidates.
Then it must be run on a fair sample of
simulated microlensing light curves to calculate its efficiency as a function
of event duration.  BEL/EB did not perform these 
steps correctly.

One problem that exists for every microlensing experiment is that whatever
signal/noise (S/N) threshold one sets, most of the events will occur close
to that threshold.   The S/N is closely related to the maximum magnification
of the event, $\Amax$, which is roughly inversely related to the 
impact parameter, $\umin$.
Since microlensing is uniformly probable in impact parameter, more events 
occur with small $\Amax$ than large $\Amax$.
In the MACHO collaboration, we experimented with over 100 statistics derived
from the light curves, investigating their usefulness using simulated
microlensing and our experience with backgrounds.  
We picked out around 20 of the most powerful statistics
in making our selection criteria and in distinguishing microlensing
from variable stars, supernovas, and noise.   We found that statistics
involving fitting the photometric
light curves with microlensing and supernova models, 
and those that used the photometric
errors were especially useful in picking out the lower S/N events. 

However, BEL/EB did not apply their neural net program to our statistics,
but instead created 5 of their own statistics, mostly based on cross
and autocorrelation of the light curve data, as well as several additional
statistics
based on colour information for supernova discrimination.
None of their statistics make direct use of the
photometric errors, and their main S/N statistic, $x_1$, does not make use
of the known microlensing light curve shape. 
Thus one would expect that they
would have difficultly identifying low magnification microlensing events,
as is evident from the MACHO events they throw out as non-microlensing.
Using less powerful statistics makes it likely that
their method would have lower efficiency at finding microlensing.
This does not have to be a problem however.  
A less efficient set of selection criteria will give a lower number
of microlensing candidates,
but, since each event is weighted by the inverse of its detection efficiency, 
if the efficiencies are properly calculated, 
the optical depth estimate should be roughly the same.
As we discuss below, BEL/EB did not do a proper efficiency calculation.

Next, for the LMC, BEL/EB ran their neural net on only 22000 out of 11.9
million light curves.  Given the expected optical depth in the range of 
$10^{-7}$ they should have expected to find less than one event.
They found 7 microlensing candidates because they also
ran the net on the 29 events that MACHO selection criteria
had picked out.  Thus their selection criteria are actually the neural
net `and'ed with MACHO selection criteria.  
Since they were sequentially applying both the MACHO selection criteria
and the neural net, it is clear that the efficiency of their method 
is necessarily lower than the MACHO selection criteria.  In order to tell
how much lower, they would need to actually apply
both the MACHO criteria and their net to a sample of simulated light curves.
They did not do this, and in fact, in Figure 3 (Evans \& Belokurov 2004b)
they show a plot of efficiency versus $\that$ that has a {\bf higher} efficiency
than our criteria 'B' efficiencies (which is our highest efficiency set).
They erroneously
state that the efficiency of our criteria 'B' is a lower limit to 
their efficiency when, in fact,
our efficiencies are an upper limit to their efficiencies since they use our
cuts before applying theirs.

We have extensive experience looking at light curves and examination
of the light curves rejected by the neural net, shows they are mostly the
low S/N events with low maximum magnification and shorter duration.  
This is what we would expect given their less powerful statistics.
Considering only the less efficient and less contaminated 
MACHO selection criteria 'A' for simplicity, 
BEL/EB keep only 3 of the 9 events with $\Amax<3$.
If this is a general rule then their real efficiency (neural net plus selection 
criteria 'A') should be lowered by roughly a factor of
$1 - (2/3)(\umin(\Amax=1.34)- \umin(\Amax=3))/\umin(\Amax=1.34) = 0.56$.
Remarkably, if one 'corrects' their
efficiencies by this factor and uses the events they select as microlensing,
one finds their estimate of optical depth increases 
to $1.1\ten{-7}$, consistent with ours (with a larger error due to
having fewer events).  Of course, this 'correction' is based upon
speculation of what the BEL/EB efficiencies are, as are the BEL/EB results.  
However, a point in favour of BEL/EB efficiencies being lower, is that
the microlensing interpretation of some of the events BEL/EB claim are 
contaminants has been strengthened by follow-up data (Bennett et al. 2004). 

As BEL/EB emphasise it is extremely important to be sure that there is
not much contamination in the final microlensing sample.
One check we did was to compare our distributions of $\umin$ with the
distribution expected from microlensing (and not expected from variable
stars or SNe).  We performed a Kolmogorov-Smirnov test and found excellent
agreement, especially after correcting for our efficiency as a function
of $\umin$.  As we pointed out in Alcock, \etal (2000), if one accepts
the high magnification events as microlensing then there {\it must}
be many more lower magnification events.  Selection criteria `A'
especially yield these events in just the right proportion.  

Another related and serious error is that if one took the BEL/EB claim of having
higher efficiencies than criteria 'B'
seriously, then in their search through the 11.9 million light curves,
they should have found some additional real microlensing that we missed.  
These events then should have been added into the optical depth estimation.
They claim that it is unlikely that they would find any additional
microlensing candidates in the 11.9 million light curves because they did
not find any in the 22000 MACHO light curve sample they investigated.  
This claim does not make sense.  
If they had found even one new event in only 22000
light curves they then should expect to find more than 500 new events in the 11.9
million.  This would give a rather large value for $\tau$.  

However, there is a serious worry that the BEL/EB net as currently configured
would, in fact, find such a large number of 
additional microlensing candidates if it were run on 11.9 million light curves.  
In BEL (2004), and EB (2004) they say they found 2 false positives in 22000 light curves
which were near the borderline between microlensing and noise events 
(and we suppose were then removed by hand?).
To get an accurate estimate of efficiencies, 
one must treat the real data in the same way as the simulated data,
so one cannot arbitrarily throw out such events.  If their false positive
rate is 2 in 22000 for random MACHO light curves, 
they would be swamped with about 1000 non-microlensing contaminates.  This would require them to tighten
their cuts and/or add more statistics, resulting in a lowering of their
efficiencies.  By relying on the MACHO selection criteria to remove
false positives from the 11.9 million light curves, 
BEL/EB were able to use less powerful statistics and looser
cuts, thus fooling themselves that their net had a high efficiency.

\subsection{Supernova identification}
The MACHO collaboration discovered the fact that SNe occurring in background
galaxies are a serious contaminant in microlensing surveys towards
the LMC, and  must be carefully identified and removed.
We use two criteria for identifying SNe.  First, using HST or
good seeing data if they exist, we check if there is
a bright background galaxy overlapping the microlensing candidate.  This
condition is very rare for a random star, and when it occurs on a microlensing
event, basically ensures the event is a SN.
Next we fit our candidates
with a SN type Ia light curve model and ask whether the SN fit is better than the
microlensing fit.  Luckily we found for our set of events that apart from
one event (event 26 which is not in set `A') these
two tests always agreed.  Thus we easily removed all supernova
contamination, especially for set `A'.  
BEL/EB did not check for background galaxies, but used colour shift
and peak symmetry information from the light curve to try to identify SNe.
They correctly identified some of our SNe, but their
SN neural net (BEL 2004) did not correctly identify 
two of these SNe (events 10 and 24).  They say they ``do not confirm the SN
classification of the Alcock, et al."
But the clear presence of a background galaxy in the images
shows that in fact these are SNe and that the BEL/EB SN net is not as reliable as
our method.  We note that event LMC-10 is one of our best SN events with a near perfect fit to our type Ia template.  BEL/EB seem to reject it because it does not show typical type Ia colours.  This illustrates the power of light curve fitting.  In the end, they leave these events out of
their final set of microlensing candidates, but did not explain why (BEL 2004; EB 2004).

\subsection{Treatment of data}
BEL/EB make the case that their neural net is superior 
to our selection criteria in picking out microlensing and not producing
false positives.  While we cannot tell for sure until they apply
it to all 11.9 million light curves and do a proper efficiency
analysis, we find this quite unlikely.

\begin{figure}
\epsfig{file=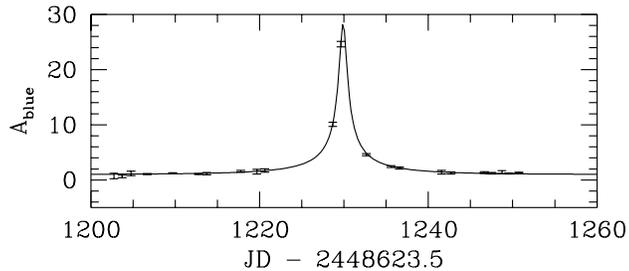,width=0.47\textwidth}
\caption{MACHO event 119.20738.3418 which, while a good microlensing candidate, was missed by the BEL/EB analysis.}
\label{fig:lc}
\end{figure}

As discussed above, BEL/EB reduce the photometric data
to a few statistics that throw away much of the information we have 
found most useful in picking out lower S/N microlensing events.
For example their statistics cannot distinguish between microlensing and any other achromactic, symmetric bump.  
In addition, they ``clean" the photometric
data by removing any point that deviates more than 3-sigma from its two
neighbours. 
Thus they remove data from any fast rising short duration peak. 
The MACHO collaboration also cleans the data by removing bad
observations, but we find these by examining all stars in a given image
and removing images that give many deviant photometric points.  
This is illustrated by event MACHO-119.20738.3418 (Figure~\ref{fig:lc}),
which is missed by BEL/EB, possibly due to their cleaning method.

In performing their
cross- and autocorrelations BEL/EB also assume that the photometric data
is evenly spaced in time, when in fact the intervals between
observations vary widely.
Therefore they may disregard information that we 
found useful and introduce unnecessary noise, especially during rapidly
rising events.  We also suspect that the simulated microlensing events
they used to train the net and to run their limited efficiency effort
are quite unlike actual microlensing events in the MACHO data.
We found that the scatter in the data and the photometric errors 
vary greatly from observation to observation,
depending strongly on seeing, local crowding, sky, etc. and
are non-Gaussian.
Thus to create
our simulated light curves we added microlensing flux to actual (unlensed) 
light curves being careful to scale the errors properly.  
If the simulated light curves
are not very similar to real microlensing then the efficiency calculation
will give incorrect results.  This is independent of whether a neural net
or more traditional selection method is used.

Finally, much of the effort in the MACHO efficiency calculation went into
understanding blending.  BEL/EB claim that this effect cancels out to first
order, and this is correct, but the 2nd order effect can
be significant.  
If one blindly uses our blending efficiencies with a different
set of selection criteria (the neural net) one is just speculating
on what the answer will be.

\section{Actual variable contamination: Event 23}
There is one variable star contaminant in
the MACHO LMC data set that seems to be well established.  The EROS collaboration (Glicenstein 2004) discovered that
event 23 bumped again after 7 years and therefore is unlikely to be
microlensing.  Since this event contributes 8\% of the optical depth
to set ``A" (6\% to set ``B"), 
all of our estimates of optical depth and Macho halo fraction should
be reduced by these amounts.    (8\% in set A and 6\% for set B).
Our estimated optical depth $\tau =  1.2 \ten{-7} $ should be reduced
to $\tau= 1.1 \ten{-7}$.
The bigger worry caused by this discovery is that there could be more such
events in our sample.  We note that the BEL/EB neural net incorrectly identified
event 23 as microlensing and so is no help here. 

In Alcock, \etal 2000,
we were worried about such contamination which is why we employed two
different sets of selection criteria.  The nature of microlensing is
such that most of the events will be at low $\Amax$ and therefore low
S/N.  We attempted to study the magnitude of possible contamination
systematic error by using a looser set of cuts (set ``B") that we felt would
be more seriously contaminated, and comparing with
a tighter set of cuts (set ``A") with
less contamination (and fewer real events).  We found that since the
efficiency of the tighter cuts was lower, the final optical depth estimates
were very similar.  We used the 17\% difference between the two methods
as an estimate of the size of the contamination error.  We note
that the 8\% reduction in $\tau$ implied by removing event 23 is within
this reported error.  Contamination will always increase the optical depth estimate, but there are other small systematic effects, such as the higher probability of missing binary events, that would shift the optical depth in the opposite direction. 

\section{Conclusion and Discussion}
We find that the MACHO collaboration estimates of optical
should be reduced by around 8\% due to the contamination of our set of events
with a variable star (event 23).  While there may be other contaminants
in our data sample, the distribution of $\umin$ and the agreement in
optical depth from our ``more contaminated" and ``less contaminated"
samples argue that this is unlikely to be substantial.
We find that the re-analysis of BEL/EB is flawed and at this time does not
make useful statements about contamination in the MACHO data set.  
We are not saying that a neural net along the lines of BEL/EB cannot be useful
in selecting microlensing events, but only that one has yet to be applied 
correctly.  Neural nets can in many cases do better than more conventional
cuts, and we applaud the effort at applying them to microlensing.  Table~\ref{tab:compare} gives a comparison of the MACHO and BEL/EB analyses.

\begin{table}
\caption{Comparison of the MACHO and BEL/EB analyses}
\label{tab:compare}
\begin{tabular}{lcc}
  \hline
  & MACHO & BEL/EB\\
  \hline
  Objects analysed & $11.9\times 10^{6}$ & $22000$\\
  Efficiency & Full & Approximated \\
  False detection rate & $<5\times 10^{-7}$ & $9\times 10^{-5}$\\
  Number of events & $13-17$ & $0(7)^\dagger$\\
  \hline
\end{tabular}
\medskip
\\
$^\dagger$ The 7 events found were found by reanalysing the MACHO events, not from the 22000 analysed.
\end{table}

It is clear that Machos cannot make up the bulk of the Milky
Way dark matter.
However,
we find that the mystery of the LMC microlensing events, 
that is, whether they indicate
a new halo population or LMC self lensing, 
is still experimentally 
unresolved, though LMC self lensing is the more popular explanation at
this time for several reasons.  Contamination by non-microlensing
events such as LMC-23 is also still a possible explanation.
Future experiments can answer this question definitively.

\medskip
We thank David Bennett, Will Sutherland, and Andrew Drake for their helpful comments.  This work supported in part by the U.S. Department of Energy under grant DEFG0390ER40546.

\label{lastpage}
\end{document}